\pdfoutput=1

\documentclass[%
 reprint,
 superscriptaddress,
 nobibnotes,
 amsmath,amssymb,
 aps,
 showkeys,
]{revtex4-2}

\usepackage{graphicx}% Include figure files
\usepackage{dcolumn}% Align table columns on decimal point
\usepackage{bm}% Bold math
\usepackage{chemformula}% Chemical formulas
\usepackage[separate-uncertainty = true]{siunitx}% SI units with separate uncertainty
\usepackage{hyperref}% Add hypertext capabilities
\hypersetup{colorlinks=true,allcolors=magenta}

\begin{document}

\title{Atomistic insights into the degradation of halide perovskites: a reactive force field molecular dynamics study}

\author{Mike Pols}
    \affiliation{Materials Simulation \& Modelling, Department of Applied Physics, Eindhoven University of Technology, 5600 MB, Eindhoven, The Netherlands}
    \affiliation{Laboratory of Inorganic Materials Chemistry, Schuit Institute of Catalysis, Department of Chemical Engineering and Chemistry, Eindhoven University of Technology, P.O. Box 513, 5600 MB, Eindhoven, The Netherlands}
    \affiliation{Center for Computational Energy Research, Department of Applied Physics, Eindhoven University of Technology, 5600 MB, Eindhoven, The Netherlands}
\author{Jos\'{e} Manuel Vicent-Luna}
    \affiliation{Materials Simulation \& Modelling, Department of Applied Physics, Eindhoven University of Technology, 5600 MB, Eindhoven, The Netherlands}
    \affiliation{Center for Computational Energy Research, Department of Applied Physics, Eindhoven University of Technology, 5600 MB, Eindhoven, The Netherlands}
\author{Ivo Filot}
    \affiliation{Laboratory of Inorganic Materials Chemistry, Schuit Institute of Catalysis, Department of Chemical Engineering and Chemistry, Eindhoven University of Technology, P.O. Box 513, 5600 MB, Eindhoven, The Netherlands}
    \affiliation{Center for Computational Energy Research, Department of Applied Physics, Eindhoven University of Technology, 5600 MB, Eindhoven, The Netherlands}
\author{Adri C.T. van Duin}
    \affiliation{Department of Mechanical Engineering, Pennsylvania State University, University Park, PA 16802, United States}
\author{Shuxia Tao}
    \email[Corresponding author: ]{s.x.tao@tue.nl}
    \affiliation{Materials Simulation \& Modelling, Department of Applied Physics, Eindhoven University of Technology, 5600 MB, Eindhoven, The Netherlands}
    \affiliation{Center for Computational Energy Research, Department of Applied Physics, Eindhoven University of Technology, 5600 MB, Eindhoven, The Netherlands}

\date{\today}

\begin{abstract}
Halide perovskites make efficient solar cells due to their exceptional optoelectronic properties, but suffer from several stability issues. The characterization of the degradation processes is challenging because of the limitations in the spatio-temporal resolution in experiments and the absence of efficient computational methods to study the reactive processes. Here, we present the first effort in developing reactive force fields for large scale molecular dynamics simulations of the phase instability and the defect-induced degradation reactions in inorganic \ch{CsPbI3}. We find that the phase transitions are driven by a combination of the anharmonicity of the perovskite lattice with the thermal entropy. At relatively low temperatures, the \ch{Cs} cations tend to move away from the preferential positions with good contacts with the surrounding metal halide framework, potentially causing its conversion to a non-perovskite phase. Our simulations of defective structures reveal that, although both iodine vacancies and interstitials are very mobile in the perovskite lattice, the vacancies have a detrimental effect on the stability, initiating the decomposition reactions of perovskites to \ch{PbI2}. Our work puts ReaxFF forward as an effective computational framework to study reactive processes in halide perovskites. 
\end{abstract}


\maketitle

\section*{Introduction}

In the past decade, halide perovskites have emerged as a promising alternative to silicon for solar cells due to their exceptional optoelectronic properties and facile fabrication methods~\cite{greenEmergencePerovskiteSolar2014, snaithPresentStatusFuture2018}. Through extensive research efforts, the efficiency of perovskite solar cells (PSCs) has risen from 3.8\% in 2009~\cite{kojimaOrganometalHalidePerovskites2009} to over 25\% in 2020~\cite{NREL2020}. Despite a considerable increase in the performance of PSCs over the years, the commercialization of perovskite solar cells is hindered by the poor long-term stability.

Halide perovskites have a three-dimensional structure with the \ch{AMX3} chemical formula, where \ch{A} is a monovalent inorganic or organic cation (\ch{Cs+}; methylammonium \ch{MA+} or formamidimium \ch{FA+}), \ch{M} is a divalent metal cation (\ch{Pb^{2+}} or \ch{Sn^{2+}}) and \ch{X} is a monovalent halide anion (\ch{I-}; \ch{Br-} or \ch{Cl-}). The metal and halide ions form a network of corner-sharing \ch{MX6} octahedra, with the centre of the cuboids formed by these octahedra occupied by the relatively large monovalent \ch{A} cation. The crystal lattice is held together by a mix of ionic and relatively weak covalent bonds, as a result which this class of materials has a soft and dynamical crystal lattice~\cite{poglitschDynamicDisorderMethylammoniumtrihalogenoplumbates1987, mashiyamaDisorderedCubicPerovskite1998, fengMechanicalPropertiesHybrid2014, sunMechanicalPropertiesOrganic2015}.

Most of the stability issues of PSCs can be traced back to the intrinsic instability of the perovskite absorber layers~\cite{niuReviewRecentProgress2015, wangStabilityPerovskiteSolar2016, correa-baenaPromisesChallengesPerovskite2017, parkIntrinsicInstabilityInorganic2019}. Such instability issues include the phase instability where it transforms to a more stable non-perovskite phase, with worse optoelectronic properties and thus a decreased power conversion efficiency of the PSC~\cite{qiuRecentAdvancesImproving2020}. Moreover, spin coating as a typical fabrication method introduces a large number of defects in the perovskite films~\cite{stranksRecombinationKineticsOrganicInorganic2014, dragutaSpatiallyNonuniformTrap2016}, the migration and accumulation of which is suggested to have a major impact on the long-term stability of perovskites and thus PSCs~\cite{carrilloIonicReactivityContacts2016, liDirectEvidenceIon2017, girolamoIonMigrationInducedAmorphization2020}. Besides, the defect-induced degradation of metal halide perovskites is often accelerated by external stimuli, such as moisture~\cite{saladoImpactMoistureEfficiencydetermining2017} and oxygen in combination with ultraviolet light~\cite{abdelmageedMechanismsLightInduced2016}.

While experimental studies offer a wide variety of insights at the macroscopic and mesoscopic scale, the interpretation of atomistic details of the degradation processes is often difficult. Computer simulations can make a significant contribution to the understanding of its atomistic and microscopic mechanisms. So far, the bulk of the computational investigations of metal halide perovskite have been done using first-principles methods based on quantum mechanics (QM)~\cite{mosconiInitioMolecularDynamics2015a, zhangInitioStaticDynamic2016, zhengUnravelingWaterDegradation2019a}. However, the large computational expense of these methods only allows for the simulation of short time scales and small system sizes.

Molecular dynamics (MD) simulations making use of classical force fields are an efficient means to study large systems at long time-scales. One of the first classical force fields for halide perovskites has been developed by Mattoni et al. for the hybrid perovskite \ch{MAPbI3}~\cite{mattoniMethylammoniumRotationalDynamics2015}. The force field has found a wide range of applications, which include the cation dynamics~\cite{mattoniMethylammoniumRotationalDynamics2015}, defect dynamics~\cite{delugasThermallyActivatedPoint2016, phungRoleGrainBoundaries2020} and dissolution in water~\cite{caddeoCollectiveMolecularMechanisms2017}. However, the potential is primarily tailored to pure perovskite systems, and therefore they can not simulate mixed perovskite compounds. Recently, some advances have been made in the development of transferable potentials for mixed perovskites, including the AMOEBA polarizable force field by Rathnayake et al. for hybrid (\ch{MAPbI3}) and inorganic (\ch{CsPbI3})~\cite{rathnayakeEvaluationAMOEBAForce2020} and a potential for \ch{CsPb(Br_{x}I_{1-x})3} by Balestra et al.~\cite{balestraEfficientModellingIon2020}. While these existing force fields have been proven to be powerful to study a wide range of dynamical properties, they can however not describe the chemical bond forming and breaking involved during the degradation of halide perovskites. From this, we conclude that a reactive force field (ReaxFF), that employs a dynamical bond order based on the interatomic distance of atomic species to describe the creation and breaking of bonds~\cite{vanduinReaxFFReactiveForce2001, senftleReaxFFReactiveForcefield2016}, can be a valuable tool to study degradation processes in halide perovskites.

Therefore, in this work we present the first effort towards the development of a ReaxFF description of halide perovskites, particularly inorganic \ch{CsPbI3}. We obtain a set of ReaxFF parameters through a training procedure against a set of accurate reference data from QM calculations. To demonstrate the applicability of our ReaxFF in investigations of dynamical and reactive processes, we perform molecular dynamics simulations to study two instability problems found in \ch{CsPbI3}: first the phase instability and then defect-accelerated decomposition of the perovskites. Combining analyses, which include a phase diagram, positional probability distributions, mean square displacements and atom trajectories, we provide important atomistic insights for both degradation mechanisms.

\section*{Results}

\subsection*{Development of a ReaxFF reactive force field for \ch{CsPbI3}} \label{sec:training}

\begin{figure*}[htbp!]
    \includegraphics[width=0.8\textwidth]{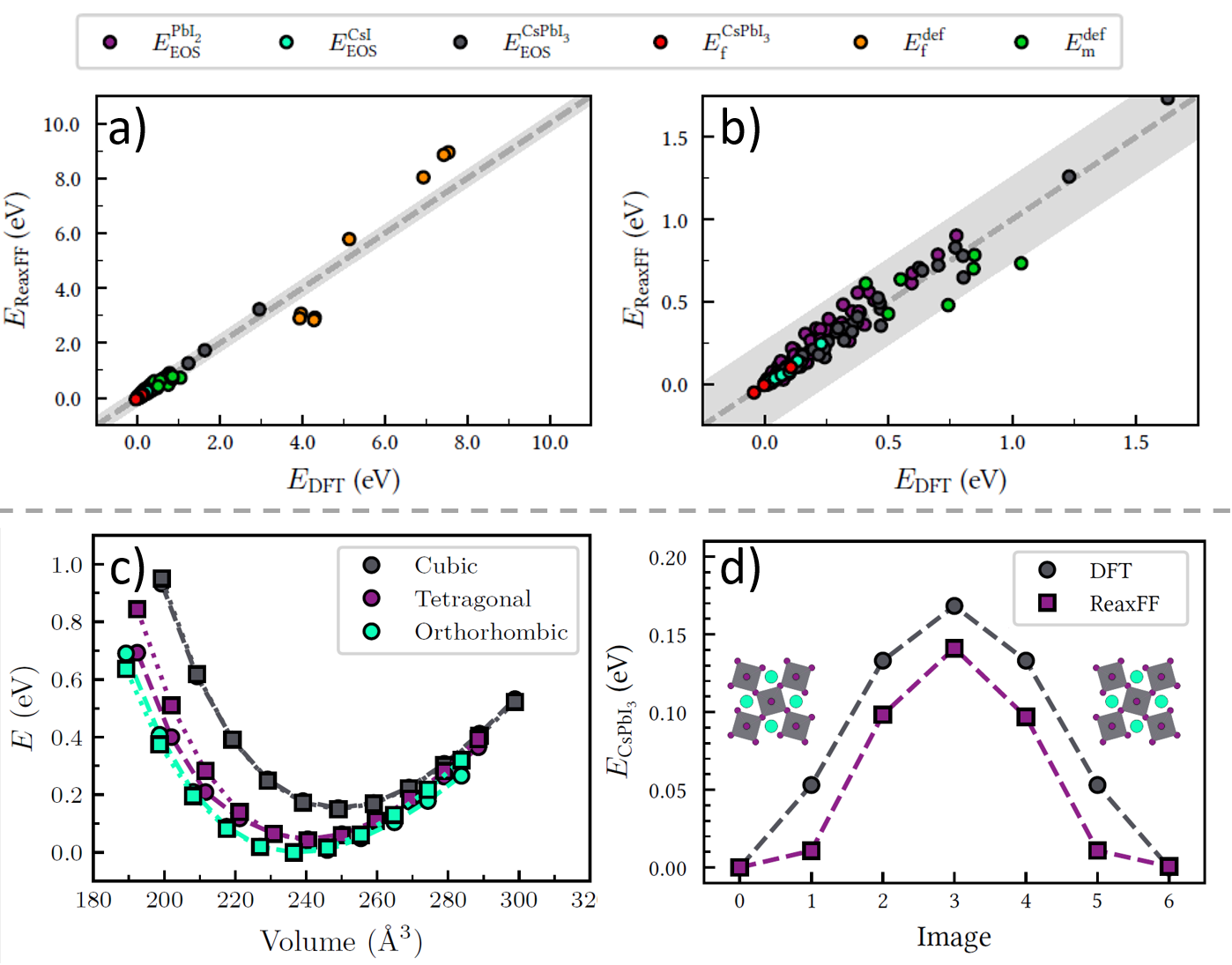}
    \caption{\label{parameter_match} \textbf{Match between \ch{I}/\ch{Pb}/\ch{Cs} ReaxFF parameters and the reference data.} \textbf{a-b} Overall agreement between the prediction with the \ch{I}/\ch{Pb}/\ch{Cs} ReaxFF parameter set and the reference data calculated with DFT calculations. \textbf{c} A comparison between the equations of state of bulk \ch{CsPbI3} from the reference data from DFT calculations (circles) and predictions with the ReaxFF parameter set (squares). \textbf{d} A comparison of the inversion barrier for the octahedral tilting pattern in tetragonal as calculated with DFT (circles) and the ReaxFF parameter set (squares).}
\end{figure*}

The ReaxFF parameters of \ch{CsPbI3} were trained against a set of reference data calculated with density functional theory (DFT). The training set included reference data for the bulk perovskite phases of \ch{CsPbI3} and its precursors \ch{CsI} and \ch{PbI2}, covering equations of state, atomic charges, formation energies, defect formation energies and defect migration barriers (see ``Methods" for details). The parameter optimization was done with a Monte Carlo-based force field (MCFF) optimizer~\cite{iypeParameterizationReactiveForce2013} as implemented in AMS2020~\cite{rugerAMS2020}. As a starting point for the parameter training, we used the previously published \ch{Cs}/\ch{I} ReaxFF parameters from the electrolyte-water parameter set published by Fedkin et al.~\cite{fedkinDevelopmentReaxFFMethodology2019}. Without any ReaxFF parameters for lead in literature, we used the atomic parameters from the parametrically similar element platinum as published by Fantauzzi et al.~\cite{fantauzziSurfaceBucklingSubsurface2015}, these were appropriately adjusted to account for the valency and atomic mass of \ch{Pb}.

Following this ReaxFF parameter optimization procedure, we obtained a ReaxFF parameter set that exhibited a good match between ReaxFF and the reference data for \ch{CsPbI3} and its precursors in our training set. The final \ch{I}/\ch{Pb}/\ch{Cs} ReaxFF parameters that result from the training procedure are found in the Supplementary Information. An overview of the agreement between the ReaxFF reactive force field and the reference data is shown in Fig. \ref{parameter_match}a) and Fig. \ref{parameter_match}b). Overall the ReaxFF parameter set shows a good agreement for the equations of state ($E_{\text{EOS}}$), perovskite formation energies ($E^{\ch{CsPbI3}}_{\text{f}}$) and defect migration barriers ($E^{\text{def}}_{\text{m}}$). To demonstrate the match between the ReaxFF parameter set and the reference data, a comparison of the bulk equations of state for \ch{CsPbI3} is shown in Fig. \ref{parameter_match}b). In regards to the defect formation energies, the ReaxFF model shows some discrepancies, with a slight over- and underestimation of the formation energies of lead and iodine vacancies, respectively. The relative magnitude of the defect formation energies is nevertheless properly captured.

\begin{table}[htbp!]
    \caption{\label{tab:geo_opt} Equilibrium geometries of the perovskite phases of \ch{CsPbI3} as calculated with the \ch{CsPbI3} ReaxFF with DFT calculations and experimental X-ray diffraction measurements as reference~\cite{marronnierAnharmonicityDisorderBlack2018}.}
    \begin{ruledtabular}
    \begin{tabular}{ccddd}
        Structure &
        Type &
        \multicolumn{1}{c}{\textrm{a (\AA)}} &
        \multicolumn{1}{c}{\textrm{b (\AA)}} &
        \multicolumn{1}{c}{\textrm{c (\AA)}} \\
        \colrule
        Cubic          & ReaxFF     & 6.29      & 6.29      & 6.29   \\
                       & DFT        & 6.29      & 6.29      & 6.29   \\
                       & Exp.       & 6.30      & 6.30      & 6.30   \\
        Tetragonal     & ReaxFF     & 8.67      & 8.67      & 6.41   \\
                       & DFT        & 8.66      & 8.66      & 6.41   \\
                       & Exp.       & 8.83      & 8.83      & 6.30   \\
        Orthorhombic   & ReaxFF     & 8.59      & 8.95      & 12.39  \\
                       & DFT        & 8.43      & 8.99      & 12.48  \\
                       & Exp.       & 8.62      & 8.85      & 12.50  \\
    \end{tabular}
    \end{ruledtabular}
\end{table}

In addition to a comparison to the reference data, we also carried out some validation tests to confirm that the obtained ReaxFF parameter set has predictive power, i.e. not only describing the entries in the training set well, but also capturing some material behaviour not explicitly trained against. Here, we carried out geometry optimizations of the different bulk perovskite phases. Starting from the DFT-optimized structure we allowed the ionic positions, cell shape and cell volume to change during the structural optimizations. The results of these calculations are shown in Table \ref{tab:geo_opt}, which shows that the ReaxFF parameter set replicates both our DFT calculations (see ``Methods" for details) and X-ray diffraction measurements from experiments well~\cite{marronnierAnharmonicityDisorderBlack2018}. We also validated the ReaxFF parameter set against a phase transition barrier not explicitly included in the training set in Fig. \ref{parameter_match}c). The energy barrier that we focus on here is the inversion barrier for the octahedral tilting in tetragonal \ch{CsPbI3}. The comparison shows that the ReaxFF calculated barrier of \SI{0.14}{\eV} agrees well with our DFT result of \SI{0.17}{\eV}.

\subsection*{Phase evolution of \ch{CsPbI3}}

\begin{figure*}[htbp!]
    \includegraphics[width=0.9\textwidth]{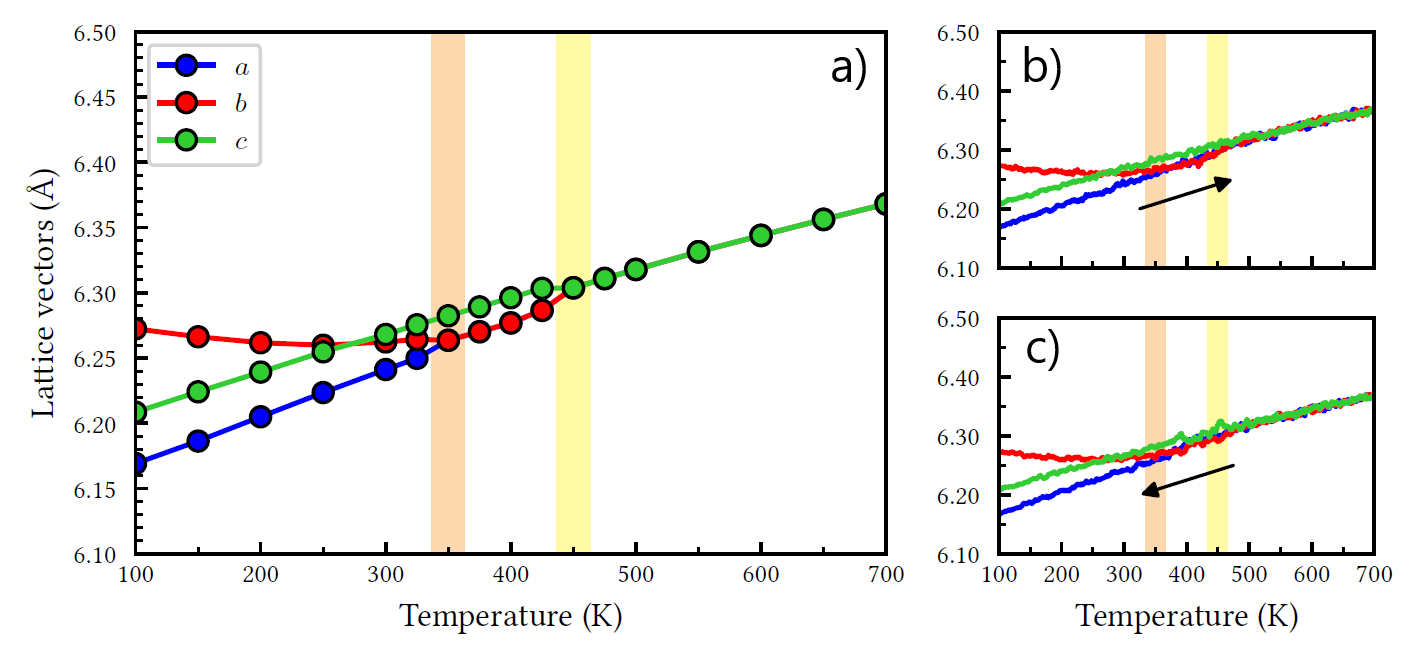}
    \caption{\label{phase_diagram} \textbf{Phase evolution of \ch{CsPbI3} between 100 K and 700 K.} \textbf{a} The temperature dependence of the lattice vectors of \ch{CsPbI3} from constant temperature simulations. \textbf{b-c} The evolution of the lattice vectors of \ch{CsPbI3} during gradual heating and cooling. The orange and yellow bars, respectively, indicate the phase transition temperatures for the orthorhombic to tetragonal and tetragonal to cubic phase transitions from the ReaxFF simulations of \ch{CsPbI3}. In all figures the pseudo-cubic lattice vectors, $a$, $b$ and $c$, of \ch{CsPbI3} are used.}
\end{figure*}

As a first application, we apply our \ch{I}/\ch{Pb}/\ch{Cs} parameter set to investigate the phase evolution of \ch{CsPbI3}. To do so, we carry out ReaxFF MD simulations for a \ch{CsPbI3} model system at a range of different temperatures between \SI{100}{\K} and \SI{700}{\K} (see ``Methods" for details). In Fig. \ref{phase_diagram}a), we show a phase diagram obtained from simulations at several discrete temperatures, for which we used a $4 \times 4 \times 4$-supercell of orthorhombic \ch{CsPbI3} (256 formal units). From the evolution of the lattice vectors in this diagram, we conclude that the \ch{CsPbI3} adopts the orthorhombic, tetragonal and cubic phase from low to progressively higher temperatures, which is in line with experimental investigations~\cite{marronnierAnharmonicityDisorderBlack2018, stoumposRenaissanceHalidePerovskites2015}. We observe that during our ReaxFF MD simulations, the \ch{CsPbI3} model system shows an orthorhombic to tetragonal phase transition at $\SI{350(10)}{\K}$ and a tetragonal to cubic phase transition at $\SI{450(10)}{\K}$. These phase transition temperatures are of a good qualitative agreement with experiments, resulting in underestimations of \SI{100}{\K} compared to experiments (\SI{457}{\K} and \SI{554}{\K})~\cite{marronnierAnharmonicityDisorderBlack2018}. We attribute these underestimations to the slight over-prediction of the lattice parameters during our simulations in comparison to experiments. Additionally, the thermal behavior of the ReaxFF MD simulations as described by the thermal expansion coefficient (\SI{12.9E-5}{\per\K}) agrees well with values from X-ray diffraction experiments (\SI{11.8E-5}{\per\K}~\cite{trotsHightemperatureStructuralEvolution2008} and \SI{15.3E-5}{\per\K}~\cite{marronnierAnharmonicityDisorderBlack2018}).

To investigate the reversibility of the phase evolution of \ch{CsPbI3}, we subjected a model perovskite system of a $6 \times 6 \times 6$-supercell of orthorhombic \ch{CsPbI3} to a continuously changing temperature. The result of these simulations is shown in Fig. \ref{phase_diagram}b) and Fig. \ref{phase_diagram}c), respectively, showing the gradual heating and cooling of the metal halide perovskite. The similarity of phase diagram obtained from gradual heating and gradual cooling simulations confirms that the phase transitions are reversible. We note here that the ReaxFF MD simulations did not always show a complete reversibility, an example of which can be seen in the Supplementary Information. We attribute this to the formation of an orthorhombic structure that consists of multiple differently oriented domains as a result of the fluctuations in the lattice during the cooling process. Such domains can be stuck with a different orthorhombic orientation and persist over time because of the lack of sufficient thermal energy. A similar phenomena, i.e. the formation of various orthorhombic domains in \ch{CsPbI3}, has also been observed in experiments~\cite{bertolottiCoherentNanotwinsDynamic2017}.

\subsection*{Phase stability of \ch{CsPbI3}}

\begin{figure*}[hbtp!]
    \includegraphics[width=0.9\textwidth]{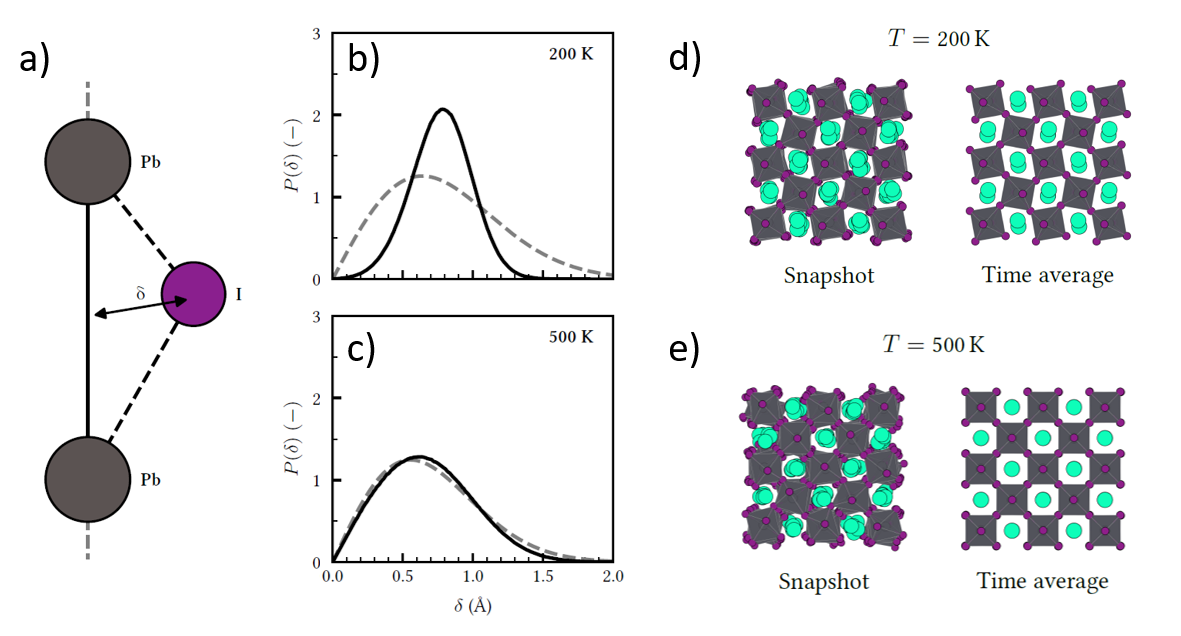}
    \caption{\label{anharmonicity} \textbf{Anharmonic character of \ch{CsPbI3} lattice.} \textbf{a} A schematic representation of the geometrical parameter $\delta$ that is used an an indicator for an anharmonic lattice. \textbf{b-c} Comparison of the probability distributions of the geometrical $\delta$ from simulations at \SI{200}{\K} and \SI{500}{\K} (solid line) against the best fit of the harmonic model (dashed line). \textbf{d-e} A comparison of the instantaneous and time-averaged structure of \ch{CsPbI3} during ReaxFF MD simulations at \SI{200}{\K} (orthorhombic phase) and \SI{500}{\K} (cubic phase).}
\end{figure*}

After having investigated the overall phase behavior of \ch{CsPbI3}, we now analyze the dynamics of the lattice at different temperatures using a method outlined by Carignano et al.~\cite{carignanoCriticalFluctuationsAnharmonicity2017} by characterizing the anharmonic character of the perovskite lattice. To do so, we define the geometrical parameter $\delta$, which is the shortest distance from an \ch{I} atom to the straight line interconnecting the two neighbouring \ch{Pb} atoms, as is shown in Fig. \ref{anharmonicity}a). A subsequent comparison of the probability distribution $P \left( \delta \right)$ from simulations against one derived for a harmonic approximation (see Supplementary Information for details) then allows for the qualitative description of the anharmonicity of the metal halide framework.

We focus on one high temperature, i.e. the cubic phase at \SI{500}{\K}, and one low temperature, i.e. orthorhombic phase at \SI{200}{\K}, in Fig. \ref{anharmonicity}b) and Fig. \ref{anharmonicity}c), respectively. By comparing the MD simulations and the best fit of the harmonic approximation, we can observe the harmonic approximation breaks down in both perovskite phases. The presence of this anharmonic character can be rationalized by Goldschmidt's principle of maximum cation-anion contact~\cite{goldschmidtCrystalStructureChemical1929, strausUnderstandingInstabilityHalide2020}. The tilting of the octahedra combined with a shift of the cation position allows for a better contact between the \ch{Cs} ions and the surrounding iodine ions and thus stabilize the perovskite structure. Interestingly, such behavior is highly sensitive to the temperature, with the largest degree of anharmonicity found at the low temperature \SI{200}{\K}, which is significantly reduced at \SI{500}{\K}. We note that this decrease in the anharmonicity is in line with the increase of the system symmetry when going from the orthorhombic to tetragonal and eventually cubic phase of \ch{CsPbI3}.

Further investigation of the simulated structures in Fig. \ref{anharmonicity}d) and Fig. \ref{anharmonicity}e), allows us to interpret the implications of the lattice anharmonicity. An inspection of the low-temperature (\SI{200}{\K}) structures reveals that the instantaneous structure from the snapshot and the time average structure closely resemble each other with just a single octahedral tilting pattern. In contrast, at the high temperature of \SI{500}{\K}, the instantaneous structure contains many local distortions, whereas the average structure is a highly symmetric one. These distinct features can be understood by taking into account the large thermal energy at elevated temperatures. At high temperatures the thermal energy is high enough to induce rapid fluctuations between many locally distorted structures. As a result of the rapid fluctuations between many differently distorted structures, the time-averaged structure becomes cubic.

\begin{figure}[hbtp!]
    \includegraphics[width=0.45\textwidth]{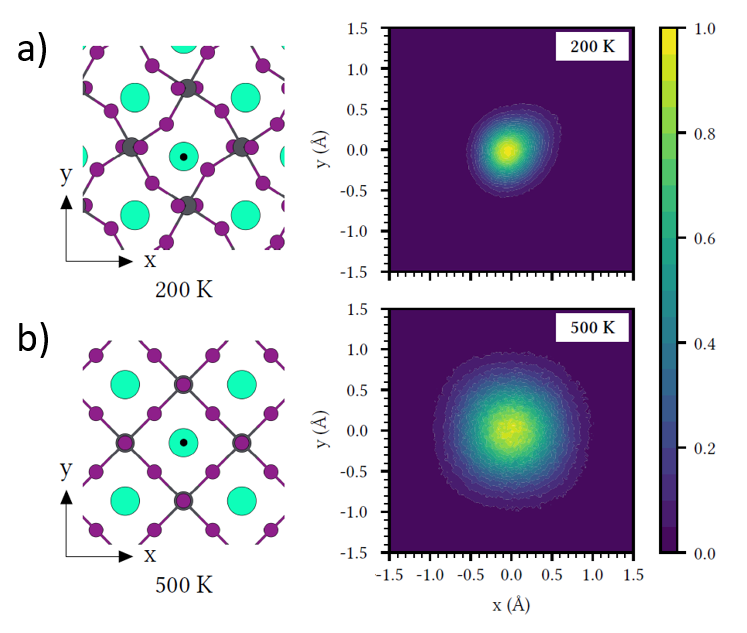}
    \caption{\label{cation_dynamics} \textbf{Dynamic structures of \ch{CsPbI3}.} \textbf{a-b} The positional probability distribution of the \ch{Cs} cations with respect to their average position in the perovskite lattice (black dot) from ReaxFF MD simulations at \SI{200}{\K} and \SI{500}{\K}.}
\end{figure}

By constructing a positional probability distribution of the \ch{Cs} cations as shown in Fig. \ref{cation_dynamics}, we next analyze the dynamics of the \ch{Cs} cations in \ch{CsPbI3}. At \SI{200}{\K}, shown in Fig. \ref{cation_dynamics}a), we observe a directed motion of the cations, with a preferential movement in the positive $xy$-direction. We attribute this directional cation movement to the anharmonicity of the perovskite crystal lattice. We rationalize that this tendency of \ch{Cs} cations to move away from the equilibrium positions can induce structural instability, potentially converting to the non-perovskite yellow phase reported in experiments~\cite{marronnierAnharmonicityDisorderBlack2018, stoumposRenaissanceHalidePerovskites2015, stoumposSemiconductingTinLead2013}. Our finding of Cs moving away from a stabilizing site to a destabilizing site is in line with experimental observations by Straus et al. from single-crystal X-ray diffraction measurements~\cite{strausUnderstandingInstabilityHalide2020}. 

In contrast, the \ch{Cs} distribution at \SI{500}{\K} in Fig. \ref{cation_dynamics}b), demonstrates that the directionality in motion of the \ch{Cs} cations is lost, resulting in an isotropic distribution. This observation can readily be explained by the lack of any long-time local structure due to the rapid fluctuations of the metal halide framework. We suggest that the fluctuations of both the \ch{Cs} cations and iodide anions result in good contacts between the two, resulting in a stabilization of the perovskite phase at high temperatures~\cite{marronnierAnharmonicityDisorderBlack2018, trotsHightemperatureStructuralEvolution2008}.

\subsection*{Ion migration in \ch{CsPbI3}}

\begin{figure*}[htbp!]
    \includegraphics[width=0.8\textwidth]{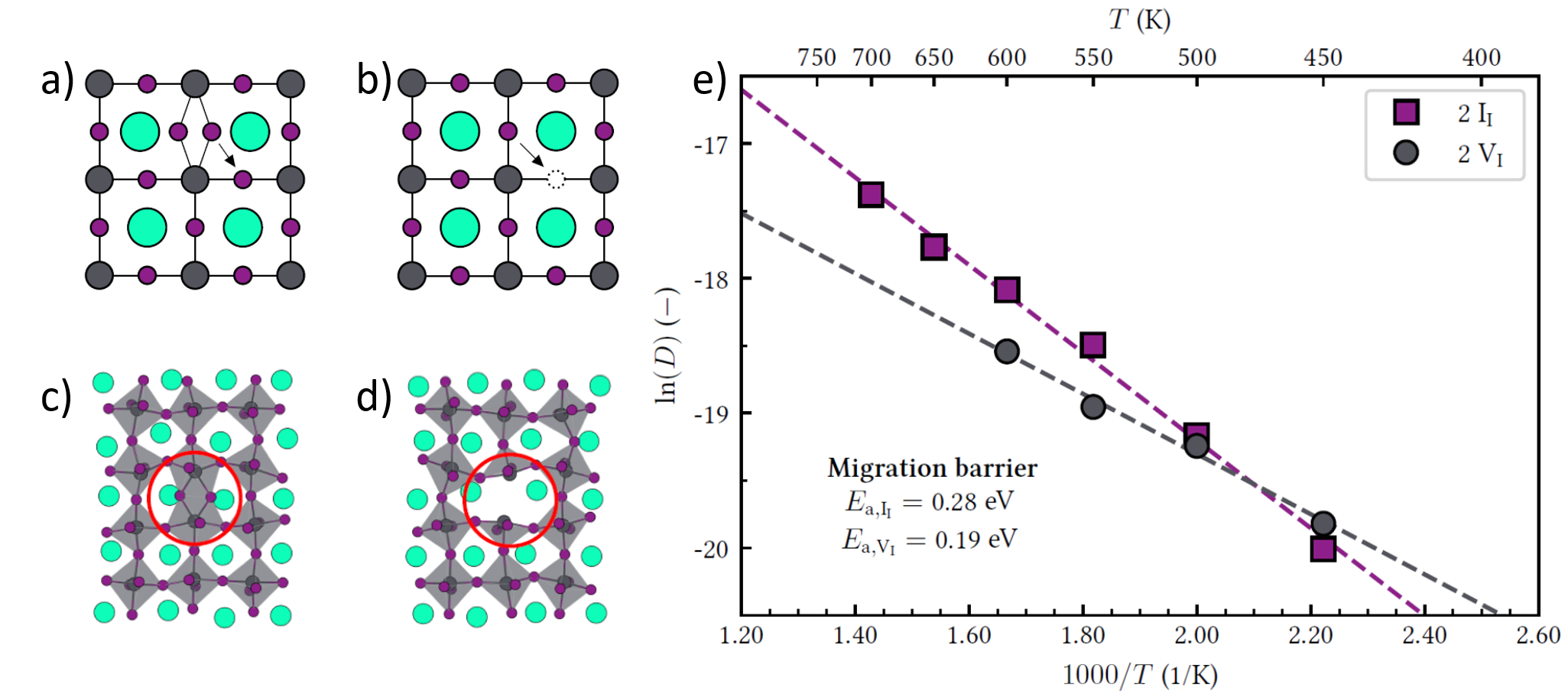}
    \caption{\label{defect_diffusion} \textbf{Defect-assisted ion migration in \ch{CsPbI3}.} \textbf{a-b} A schematic overview of the interstitial-assisted and vacancy-assisted iodine migration mechanism in \ch{CsPbI3}, respectively. \textbf{c-d} Defect geometries of, respectively, an iodine interstitial and vacancy point defect from ReaxFF simulation snapshots of \ch{CsPbI3} at \SI{500}{\K}. \textbf{e} The temperature evolution of the self-diffusion of iodine point defects as obtained from ReaxFF simulations of \ch{CsPbI3}.}
\end{figure*}

As mentioned earlier, the stability of metal halide perovskites and thus perovskite solar cells is impacted by the migration of ions. For \ch{MAPbI3}, two halide migration mechanisms have been proposed to be the major contributor to ion migration in the hybrid metal halide perovskite~\cite{azpirozDefectMigrationMethylammonium2015, delugasThermallyActivatedPoint2016, yangFastSelfdiffusionIons2016}. Here, we employ our \ch{I}/\ch{Pb}/\ch{Cs} ReaxFF parameters to investigate the relative importance of the two ion migration mechanisms, an interstitial-assisted mechanism in Fig. \ref{defect_diffusion}a) and vacancy-assisted mechanism in Fig. \ref{defect_diffusion}b), in inorganic \ch{CsPbI3}.

To do so, we carry out molecular dynamics simulations with our ReaxFF parameter set to probe the temperature evolution of the self-diffusion coefficients of the aforementioned two types of iodine point defects in inorganic \ch{CsPbI3}. The model perovskite system used during these simulations is a $4 \times 4 \times 4$-supercell of orthorhombic \ch{CsPbI3}. For both, the vacancies and interstitials, two defects were created and spaced at least \SI{20}{\angstrom} apart to ensure a homogeneous spread of the defects. As a result of this, the defective bulk perovskite structures had a defect concentration of \SI{3E19}{\per\cubic\cm} at \SI{500}{\K}. The ReaxFF simulations are done at atmospheric pressure at temperatures ranging from \SI{450}{\K} to \SI{700}{\K}, from which the self-diffusion coefficients are calculated (see ``Methods" for details). The defect geometries obtained from ReaxFF simulations are shown in Fig. \ref{defect_diffusion}c) and \ref{defect_diffusion}d). 

The temperature evolution of the self-diffusion of both types of iodine point defects is shown in Fig. \ref{defect_diffusion}e) with a complete overview of the diffusion coefficients given in the Supplementary Information. Focusing on the self-diffusion coefficients, it shows that in the investigated temperature range both types of defect exhibit similar rates of diffusion. Near the low end of the temperature range (\SI{450}{\K} - \SI{500}{\K}) similar diffusion coefficients are found. At higher temperatures ($> \SI{600}{\K}$) we find that the interstitials show a rate of diffusion that is twice as high as that of the vacancies.

An analysis of diffusion coefficients shows that the temperature dependence of the self-diffusion coefficients is well-described by a single Arrhenius relation. As a result of this, we associate the migration of both point defects with a single activation energy for the investigated temperature range. The migration barrier for the iodine interstitials was determined at $E_{\text{a,I}_{\ch{I}}} = \SI{0.28}{\eV}$ and for the iodine vacancies at $E_{\text{a,V}_{\ch{I}}} = \SI{0.19}{\eV}$, with a prefactor of $D_{\text{0,I}_{\ch{I}}} = \SI{3.1e-6}{\square\cm\per\s}$ and $D_{\text{0,V}_{\ch{I}}} = \SI{3.6e-7}{\square\cm\per\s}$. The relatively low energy barriers for defect migration mechanisms indicate that both ion migration processes readily occur in \ch{CsPbI3}. Specifically, our value for the energy barrier of iodine vacancy migration (\SI{0.19}{\eV}) matches well with those observed for halogen vacancies in the inorganic perovskites \ch{CsPbBr3} (\SI{0.25}{\eV}) and \ch{CsPbCl3} (\SI{0.29}{\eV}) as measured by Mizusaki et al. using impedance spectroscopy~\cite{mizusakiIonicConductionPerovskitetype1983}.

\subsection*{Defect-accelerated degradation of \ch{CsPbI3}} \label{sec:defect_degradation}

\begin{figure*}[htbp!]
    \includegraphics[width=0.8\textwidth]{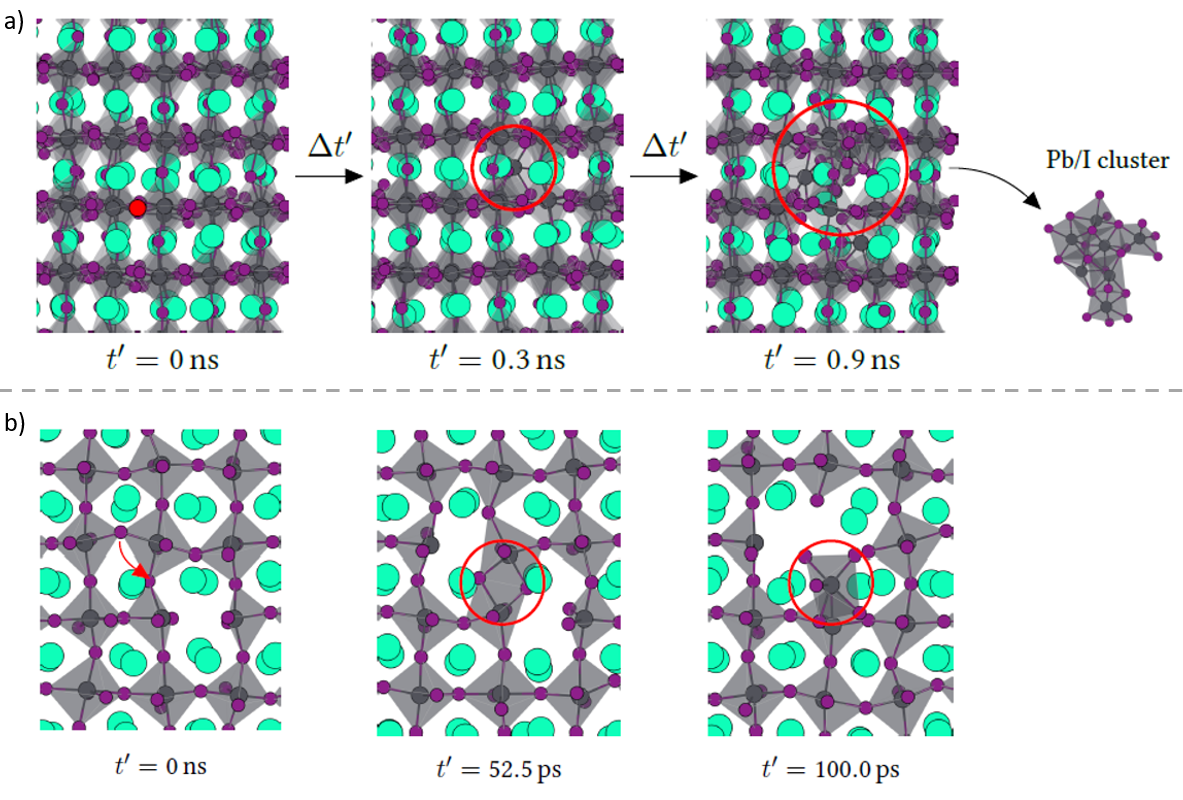}
    \caption{\label{defect_degradation} \textbf{Defect-accelerated perovskite decomposition in \ch{CsPbI3}.} \textbf{a} An overview of the different steps in the perovskite decomposition process of \ch{CsPbI3} exhibited in the presence of an iodine vacancy (red dot) at \SI{600}{\K}. The perovskite decomposes by forming a \ch{Pb}/\ch{I} cluster near the iodine vacancy. \textbf{b} Snapshots from the initial stages of the perovskite decomposition mechanism, which shows the formation of an iodine Frenkel defect that results in a small complex of edge-sharing metal halide octahedra that can easily break away from the lattice to form a \ch{Pb_{x}I_{y}} complex.}
\end{figure*}

Finally, we observe that at high temperatures the vacancy-rich \ch{CsPbI3} systems have the tendency to result in the decomposition of the perovskite structure. In Fig. \ref{defect_degradation}a) we show snapshots of a ReaxFF simulation of \ch{CsPbI3} with two iodine vacancies at \SI{600}{\K} that exhibits structural decomposition. At the onset of the simulation, the iodine vacancies are dispersed in the perovskite phase. After some time, we observe that a lead species close to the iodine vacancy moves away from its position in the lattice, forming a locally \ch{Pb}/\ch{I}-rich domain. Further evolution of the system causes the \ch{Pb}/\ch{I}-rich region to grow in size, resulting in the formation of a \ch{Pb}/\ch{I} cluster, the structure of which is highlighted in the figure.

To elucidate the details of the degradation mechanism of the metal halide perovskite, a more detailed overview of the initial stages of the perovskite decomposition is shown in Fig. \ref{defect_degradation}b). The snapshots show that the degradation process begins with the formation of an iodine Frenkel defect close to the existing iodine vacancy; one iodine atom leaves its original position to create a vacancy and at the same time forms one iodine interstitial site. As a result of this, two octahedra are connected by the newly formed interstitial site, forming a complex of edge-sharing octahedra, causing the \ch{PbI_{x}} octahedra in this complex to only be weakly bound to the rest of the perovskite lattice. Consequently, the lead species of either of these octahedra easily move away from their position in the lattice, forming a face-sharing \ch{Pb_{x}I_{y}} complex, which subsequently grows to a larger \ch{Pb}/\ch{I} cluster.

We hypothesize that such face-sharing \ch{Pb_{x}I_{y}} complexes serve as the nucleation centre for the decomposition of the perovskite structure, eventually leading to the decomposition of the metal halide perovskite into \ch{PbI2}. Our finding corroborates with recent observations from transmission electron microscopy experiments by Manekkathodi et al., in which \ch{PbI2} nanoparticles are detected in the vicinity of lattice defects such a grain boundaries in mixed metal halide perovskites~\cite{manekkathodiObservationStructuralPhase2020}. Moreover, our finding provides an atomistic interpretation of the fact that iodine-rich conditions can have a stabilizing effect on metal halide perovskites~\cite{kyeVacancyDrivenStabilizationCubic2019}, by inhibiting the formation of iodine vacancies that, from our ReaxFF simulations, appear to accelerate the degradation of the perovskites.

\section*{Discussion}

In summary, we present a first reactive molecular dynamics study of metal halide perovskites using ReaxFF, using \ch{CsPbI3} as an example. A Monte Carlo-based optimization algorithm is used to obtain a set of ReaxFF parameters, by training against a set of accurate quantum mechanical reference data from DFT calculations. Through a set of validation tests, we confirm that our ReaxFF parameter set has ample predictive power.

Using molecular dynamics simulations, we demonstrate that the transition between the different phases of \ch{CsPbI3}, i.e. from  orthorhombic to tetragonal and to cubic phases, are results of a combination of the anharmonic nature of the perovskite lattice and the thermal entropy. Additionally, we explain the phase instability by coupling the dynamics of the \ch{Cs} cations with the anharmonicity of the perovskite lattice. We suggest that the \ch{Cs} cations prefer to locate at positions for good contact with the metal halide framework, which is facilitated by rapid dynamical fluctuations at high temperatures. However, at relatively low temperatures, \ch{Cs} cations tend to move away from this preferential position, resulting in an instability that potentially causes the conversion of the perovskite phase to the non-perovskite phase. As we pointed out earlier, the fundamental reason for such instability is the small size of the \ch{Cs} compared to the metal halide framework, therefore mixing with larger organic cations~\cite{zhangImprovedPhaseStability2019, liuInsightImprovedPhase2020} or smaller anions such as the \ch{Br} anion~\cite{suttonBandgapTunableCesiumLead2016} can mitigate the observed structural distortion and improve the phase stability of CsPbI3.

Our ReaxFF simulations of defect-rich \ch{CsPbI3} reveal that both interstitial-assisted and vacancy-assisted migration play a substantial role in the migration of ions in \ch{CsPbI3}. We find that the iodine vacancies are detrimental to the stability of metal halide perovskites, by facilitating the formation of iodine Frenkel defects in crystal lattice, which eventually grow into \ch{Pb}/\ch{I} clusters, resulting in the decomposition of the perovskite lattice. Our findings suggest that materials engineering strategies that reduce or passivate the concentration of vacancy defects are important to improve the stability of halide perovskites. Such strategies include but are not limited to the synthesis of halide perovskites in \ch{I}-rich conditions~\cite{eperonInorganicCaesiumLead2015}, the inclusion of additives~\cite{guoAdditiveEngineeringHighPerformance2020}.

Our work paves the way for large-scale reactive molecular dynamics simulations of metal halide perovskites. We expect that the set of ReaxFF parameters presented in this work can readily be expanded to cover a broader range of metal halide perovskite compositions and their interactions with contact layers in solar cells. These future developments in new force fields will allow the study of several other reactive processes for realistic compositions that is relevant to large-scale applications in perovskite optoelectronics.

\section*{Methods} \label{sec:methods}

\subsection*{Generation of training data}

The reference data in our training set was generated using the VASP software package~\cite{kresseInitioMolecularDynamics1993, kresseInitioMoleculardynamicsSimulation1994, kresseEfficiencyAbinitioTotal1996, kresseEfficientIterativeSchemes1996}. Following extensive exchange-correlation (XC) functional tests (see Supplementary Information), all reference data was calculated using the PBE exchange-correlation functional~\cite{perdewGeneralizedGradientApproximation1996, perdewGeneralizedGradientApproximation1997}, with the long-range dispersive interactions being accounted for by the DFT-D3(BJ) dispersion correction~\cite{grimmeConsistentAccurateInitio2010a, grimmeEffectDampingFunction2011}. The outermost electrons of \ch{Cs} (5s\textsuperscript{2}5p\textsuperscript{6}6s\textsuperscript{1}); \ch{Pb} (5d\textsuperscript{10}6s\textsuperscript{2}6p\textsuperscript{2}) and \ch{I} (5s\textsuperscript{2}5p\textsuperscript{5}) were treated as valence electrons, the electron-ion interaction was modeled with the projector-augmented wave (PAW) method~\cite{blochlProjectorAugmentedwaveMethod1994, kresseUltrasoftPseudopotentialsProjector1999}. Furthermore, the plane-wave basis set was expanded to an energy cutoff of \SI{500}{\eV} with a Brillouin zone integration using Monkhorst-Pack meshes~\cite{monkhorstSpecialPointsBrillouinzone1976}.

The equilibrium geometry of all materials was obtained from structural relaxations. During these relaxations we allowed the ionic positions, cell shape and cell volume to change until the energy and force converged to within \SI{1E-3}{\meV} and \SI{10}{\meV\per\angstrom}, respectively. Here we made use of the following $k$-space grids, which resulted in an energy convergence to within \SI{1}{\meV}/atom: \ch{PbI2}: $11 \times 11 \times 7$; \ch{CsI}: $12 \times 12 \times 12$; cubic \ch{CsPbI3}: $10 \times 10 \times 10$; tetragonal \ch{CsPbI3}: $7 \times 7 \times 10$; orthorhombic \ch{CsPbI3}: $7 \times 7 \times 5$; yellow phase \ch{CsPbI3}: $13 \times 6 \times 4$. The atomic charges were calculated for these equilibrium geometries with the Bader charge analysis method~\cite{henkelmanFastRobustAlgorithm2006, sanvilleImprovedGridbasedAlgorithm2007, tangGridbasedBaderAnalysis2009, yuAccurateEfficientAlgorithm2011}. Whenever a monolayer material was modeled (e.g. \ch{PbI2} monolayer), we employed a vacuum layer of at least \SI{15}{\angstrom} to prevent interactions between the periodic images of the monolayer. The equations of state were generated by straining the lattice vectors of the equilibrium geometries, and subsequently allowing the ionic positions to relax to the above-mentioned energy and force convergence criteria.

Defect calculations were done in both \ch{PbI2} monolayers and the orthorhombic phase of \ch{CsPbI3}. To limit the interactions between the periodic images of the defects, these calculations employed supercell geometries. For \ch{PbI2} and \ch{CsPbI3}, respectively, a $4 \times 4 \times 1$ and $2 \times 2 \times 1$ supercell was used with the $k$-points scaled to $3 \times 3 \times 1$ and $2 \times 2 \times 3$. The computational cost of these defect calculations was reduced by making the convergence criteria less strict, for \ch{PbI2} this resulted in an energy and force convergence criterion of \SI{1E-1}{\meV} and \SI{30}{\meV\per\angstrom}, whereas for \ch{CsPbI3} it was set to \SI{1E-2}{\meV} and \SI{50}{\meV\per\angstrom}. The defect formation energies were determined from the difference in energy between the defective and corresponding pristine structures. The defect migration barriers were determined from transition state calculations with five intermediate geometries, using the Climbing Image Nudged Elastic Band (CI-NEB) method~\cite{henkelmanImprovedTangentEstimate2000, henkelmanClimbingImageNudged2000}.

\subsection*{Molecular dynamics}

All of the ReaxFF MD simulations in this work were carried out in AMS2020~\cite{rugerAMS2020}. Prior to the MD simulations, all system geometries were optimized with the \ch{CsPbI3} ReaxFF parameter set. For all simulations we employed a simulation timestep of \SI{0.25}{\fs} and damping constants of $\tau_{T} = \SI{100}{\fs}$ and $\tau_{p} = \SI{2500}{\fs}$ for thermostat and barostat, respectively. A chain length of 10 was used each time a Nos\'{e}-Hoover chains (NHC) thermostat was employed.

In the simulation of the perovskite phase diagram we first equilibrated the model system to its target temperature and pressure in an NPT-ensemble. During this equilibration stage of \SI{50}{\ps} we employed a Berendsen thermostat and Berendsen barostat~\cite{berendsenMolecularDynamicsCoupling1984}. For both approaches, the production runs were started from the positions and velocities of the final frame of the simulation run. The production runs were carried out in an NPT-ensemble, where the conditions were controlled by a NHC-thermostat~\cite{martynaNoseHooverChains1992} and MTK-barostat~\cite{martynaConstantPressureMolecular1994}. In the constant temperature approach, the model system was kept to a constant temperature for a duration of \SI{0.5}{\ns}, where the values of the lattice vectors were determined with the method outlined in the Supplementary Information. For the continuous heating or cooling approach to the phase diagram, the temperature of the model system was continuously varied through a linear temperature control over the NHC-thermostat. Here, we varied the thermostat temperature with a constant rate of change of $\frac{d T}{d t} = \SI{5E-4}{\K\per\fs}$, which results in a temperature change of \SI{600}{\K} over a simulation time of \SI{1.2}{\ns}. The lattice vectors during the heating and cooling simulations were averaged using a running average of \SI{10}{\ps}.

For the investigation of the lattice and ion dynamics of the metal halide perovskite, we employed similar simulation stages and settings as used during the constant temperature simulations. However, to increase the statistics of the simulations, the simulation time during the production runs was increased to \SI{2}{\ns}. In the construction of the positional probability distribution of the \ch{Cs} cations we only used the equivalent cations that were located at similar sites in the perovskite lattice. A model system of 256 formal units of \ch{CsPbI3}, contains four groups of 64 equivalent \ch{Cs} cations.

During the defect simulations, each defective bulk system was equilibrated to the target temperature and pressure using a two-step equilibration process in an NPT-ensemble. During the first equilibration step, we made use of a Berendsen thermostat and Berendsen barostat, with the second equilibration step employing a NHC-thermostat and MTK-barostat. Equilibration times of \SI{50}{\ps} and \SI{100}{\ps} were used for each equilibration step for the systems with iodine vacancies and iodine interstitials, respectively. Once equilibrated, the production run of each system was started from the final frame of the equilibration process. The production runs were carried out an NVT-ensemble in which the temperature was controlled by an NHC-thermostat. The total simulation time of these production runs was \SI{2.5}{\ns} for the iodine vacancies and \SI{2.0}{\ns} for the systems with iodine interstitials.

\subsection*{Diffusion coefficients}

The self-diffusion coefficients of the point defects were obtained from the atom trajectories using the Einstein approach to the diffusion coefficient~\cite{einsteinUberMolekularkinetischenTheorie1905}. In this method, the self-diffusion coefficient $D$ of a species is obtained from the average displacement of those species over time, which is described by the mean square displacement (MSD), which we define as
\begin{equation}
    \text{MSD} \left( t \right) = \left\langle \frac{1}{N} \sum^{N}_{i = 1} \lvert \vec{r}_{i} \left( t \right) - \vec{r}_{i} \left( 0 \right) \rvert^{2} \right\rangle_{t_{0}},
\end{equation}
where $N$ is the number of particles of the species of interest, $\vec{r}_{i} \left( t \right)$ describes the particle position after a time $t$, $\vec{r}_{i} \left( 0 \right)$ the starting position of the particle and $\langle \cdots \rangle_{t_{0}}$ indicating the averaging over different time origins $t_{0}$. The diffusion coefficient can be calculated from the MSD as
\begin{equation}
    D = \frac{1}{2 d} \lim_{t \to \infty} \frac{d}{dt} \text{MSD} \left( t \right),
\end{equation}
where $d$ is the number of dimensions in which the species can move ($d = 1,2 \text{ or } 3$). We evaluated the mean square displacements of the species on the time interval from \SI{0.5}{\ns} to \SI{2.0}{\ns} and from \SI{0.5}{\ns} to \SI{1.5}{\ns} for the iodine interstitials and iodine vacancies, respectively. 

To analyze the temperature evolution of the self-diffusion coefficients of the point defects, we used the Arrhenius relation~\cite{arrheniusUberDissociationswarmeUnd1889}. For diffusive processes in solids, this Arrhenius relation is of the following form
\begin{equation}
    D = D_{0} \cdot \exp \left( - \frac{E_{\text{a}}}{k_{\text{B}} T} \right),
\end{equation}
in which $D_{0}$ is the prefactor, $E_{\text{a}}$ the energy barrier for the defect migration, $k_{\text{B}}$ the Boltzmann constant and $T$ the temperature.

\section*{Data Availability}

All data generated and analyzed during this study are available from the corresponding author upon reasonable request.

\section*{Acknowledgements}
M.P. and S.T. acknowledge funding by the Computational Sciences for Energy Research (CSER) tenure track program of Shell and NWO (Project No. 15CST04-2); J.M.V.L. and S.T. acknowledge NWO START-UP from the Netherlands.

\section*{Author contributions}

The project was conceived and planned by S.T. All calculations were done by M.P. and guided by J.M.V.L., I.F., A.C.T.v.D. and S.T. The first version of the manuscript was written by M.P. All authors contributed to the interpretation of the results and to the final version of the manuscript.

\section*{Competing interests}

The authors declare no competing interests.

\bibliography{ms}% Produces the bibliography via BibTeX.

\end{document}

\title{Supplementary Information \\ Atomistic insights into the degradation of halide perovskites: a reactive force field molecular dynamics study}

\author{Mike Pols}
    \affiliation{Materials Simulation \& Modelling, Department of Applied Physics, Eindhoven University of Technology, 5600 MB, Eindhoven, The Netherlands}
    \affiliation{Laboratory of Inorganic Materials Chemistry, Schuit Institute of Catalysis, Department of Chemical Engineering and Chemistry, Eindhoven University of Technology, P.O. Box 513, 5600 MB, Eindhoven, The Netherlands}
    \affiliation{Center for Computational Energy Research, Department of Applied Physics, Eindhoven University of Technology, 5600 MB, Eindhoven, The Netherlands}
\author{Jos\'{e} Manuel Vicent-Luna}
    \affiliation{Materials Simulation \& Modelling, Department of Applied Physics, Eindhoven University of Technology, 5600 MB, Eindhoven, The Netherlands}
    \affiliation{Center for Computational Energy Research, Department of Applied Physics, Eindhoven University of Technology, 5600 MB, Eindhoven, The Netherlands}
\author{Ivo Filot}
    \affiliation{Laboratory of Inorganic Materials Chemistry, Schuit Institute of Catalysis, Department of Chemical Engineering and Chemistry, Eindhoven University of Technology, P.O. Box 513, 5600 MB, Eindhoven, The Netherlands}
    \affiliation{Center for Computational Energy Research, Department of Applied Physics, Eindhoven University of Technology, 5600 MB, Eindhoven, The Netherlands}
\author{Adri C.T. van Duin}
    \affiliation{Department of Mechanical Engineering, Pennsylvania State University, University Park, PA 16802, United States}
\author{Shuxia Tao}
    \email[Corresponding author: ]{s.x.tao@tue.nl}
    \affiliation{Materials Simulation \& Modelling, Department of Applied Physics, Eindhoven University of Technology, 5600 MB, Eindhoven, The Netherlands}
    \affiliation{Center for Computational Energy Research, Department of Applied Physics, Eindhoven University of Technology, 5600 MB, Eindhoven, The Netherlands}

\date{\today}

\maketitle

\tableofcontents

\clearpage

\section*{Supplementary Notes}

\subsection*{Supplementary Note 1: Benchmark of density functional theory settings}

To assess the quality of different levels of theory within the framework of density functional theory (DFT), we tested a variety of methods as implemented in the Vienna Ab-Initio Simulation Package (VASP)~\cite{kresseInitioMolecularDynamics1993, kresseInitioMoleculardynamicsSimulation1994, kresseEfficiencyAbinitioTotal1996, kresseEfficientIterativeSchemes1996}. Noting that we aim to obtain a ReaxFF reactive force field that can accurately describe the phase and defect behavior of the \ch{CsPbI3} metal halide perovskite, we focused in particular on the accuracy of the computational methods for the experimentally observed bulk phases of \ch{CsPbI3} and its precursors (\ch{CsI} and \ch{PbI2}). Reminding ourselves of the extensive amount of data that is typically required for a training set for an empirical force field, we limit our tests to the computationally efficient local and semi-local functionals: local density approximation (LDA), generalized gradient approximation (GGA) and meta-GGA. The exchange-correlation functionals that we tested include LDA~\cite{ceperleyGroundStateElectron1980}, PBE~\cite{perdewGeneralizedGradientApproximation1996, perdewGeneralizedGradientApproximation1997}, PBEsol~\cite{perdewRestoringDensityGradientExpansion2008, perdewErratumRestoringDensityGradient2009} and SCAN~\cite{sunStronglyConstrainedAppropriately2015}. Moreover, our tests also covered a range of methods that explicitly include dispersive interactions into the calculations, these methods include DFT-D3~\cite{grimmeConsistentAccurateInitio2010}, DFT-D3(BJ)~\cite{grimmeEffectDampingFunction2011} and rVV10~\cite{pengVersatileVanWaals2016}. For the dispersive interactions in combination with the SCAN functional, we used parameters from literature for the DFT-D3/DFT-D3(BJ)~\cite{brandenburgBenchmarkTestsStrongly2016} and rVV10~\cite{pengVersatileVanWaals2016} dispersive interactions.

\begin{figure}[htbp!]
    \includegraphics[width=0.9\textwidth]{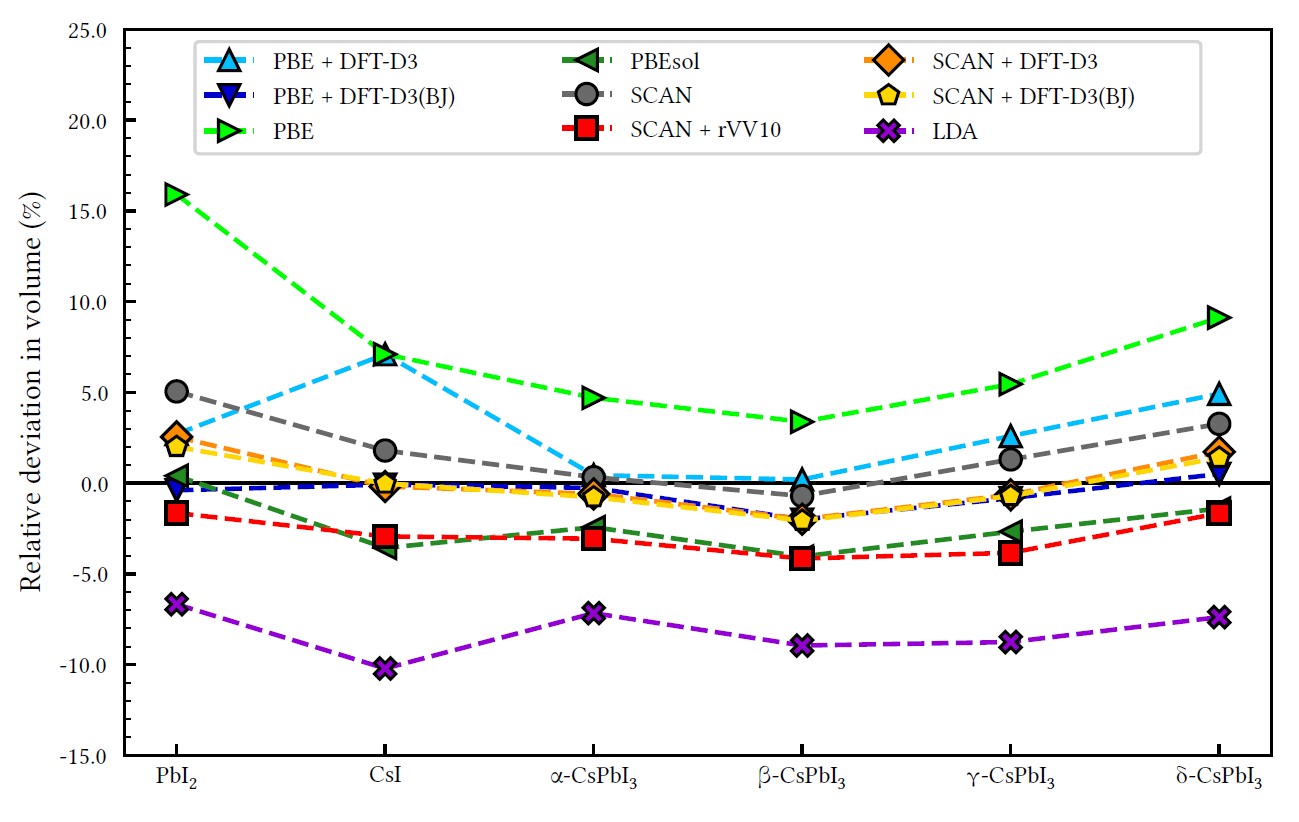}
    \caption{\label{volume_comparison} \textbf{Volume benchmark of a range of DFT settings.} Unit cell volumes for a variety of density functional approximations compared to experimentally determined unit cell volumes. With \textalpha-, \textbeta-, \textgamma-, \textdelta-\ch{CsPbI3} denoting the cubic, tetragonal, orthorhombic and yellow phase of \ch{CsPbI3}, respectively.}
\end{figure}

\begin{figure}[htbp!]
    \includegraphics[width=0.9\textwidth]{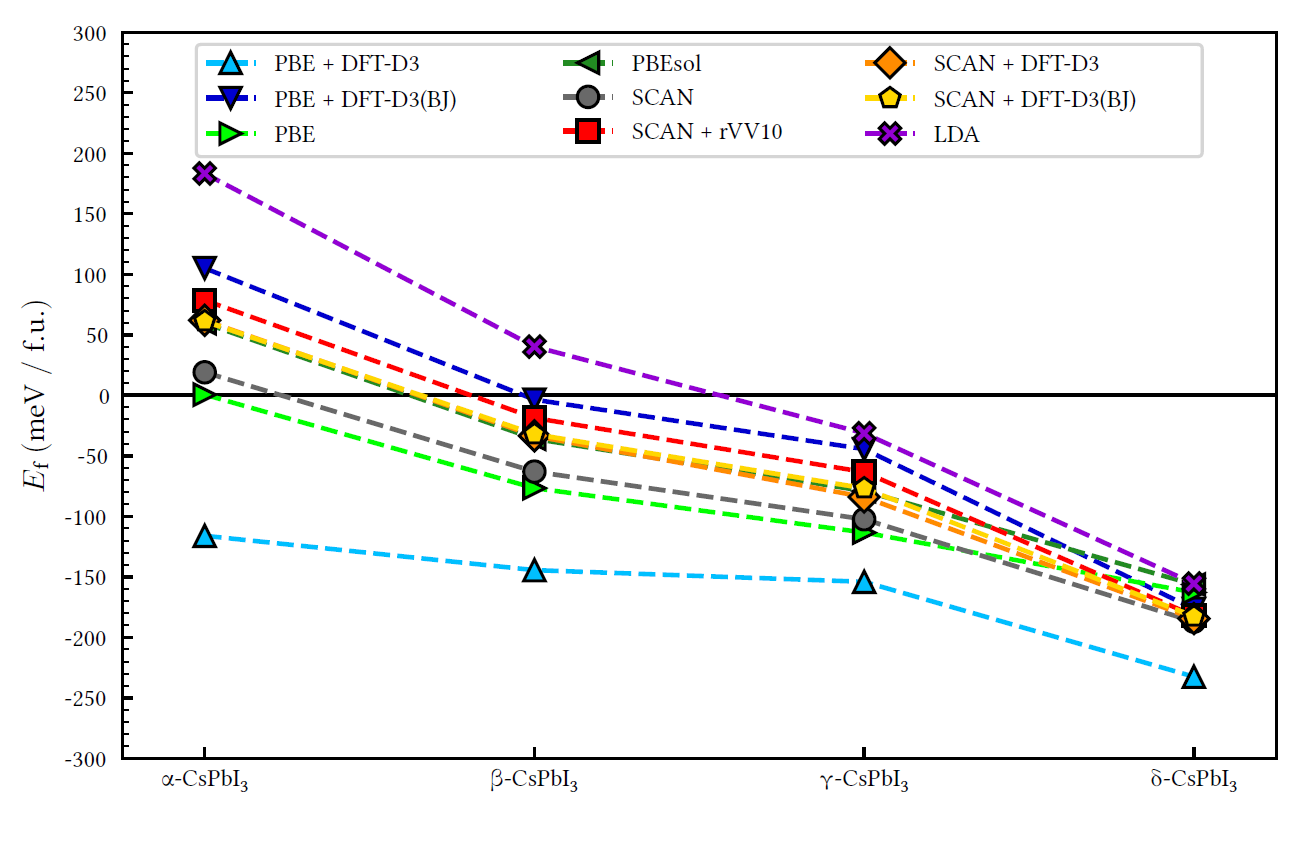}
    \caption{\label{energy_comparison} \textbf{Phase stability benchmark of a range of DFT settings.} The formation energies of the different material phases of \ch{CsPbI3} per formal unit as calculated with different density functional approximations. With \textalpha-, \textbeta-, \textgamma-, \textdelta-\ch{CsPbI3} denoting the cubic, tetragonal, orthorhombic and yellow phase of \ch{CsPbI3}, respectively.}
\end{figure}

To benchmark the accuracy of the different density functional approximations, we looked at the material geometries and relative stability of the different perovskite phases. All structures were fully relaxed according to the calculation settings for the equations of state as found in the ``Methods'' section. In the geometry comparison we referenced a variety of material phases, including \ch{PbI2}~\cite{flahautCrystallization2H4H2006}, \ch{CsI}~\cite{rymerLatticeConstantCaesium1951}, cubic, tetragonal and orthorhombic \ch{CsPbI3}~\cite{marronnierAnharmonicityDisorderBlack2018} and yellow phase \ch{CsPbI3}~\cite{stoumposSemiconductingTinLead2013}, to their experimentally reported geometries. Here we used the relative change in unit cell volume with respect to the experimental unit cell as a measure of the accuracy of the calculations. An overview of the results is found in Fig. \ref{volume_comparison}. The phase stability was determined from the formation energies of the perovskites. We define the formation energy per formal unit of \ch{CsPbI3} as
\begin{equation}
    E_{\mathrm{f}} = E_{\ch{CsPbI3}} - E_{\ch{CsI}} - E_{\ch{PbI2}},
\end{equation}
where $E_{\ch{CsPbI3}}$, $E_{\ch{CsI}}$ and $E_{\ch{PbI2}}$ are, respectively, the energies per formal unit of \ch{CsPbI3}, \ch{CsI} and \ch{PbI2}. Not only did this formation energy give us a measure of the stability of the different phases of \ch{CsPbI3} with respect to its precursors, it also allowed for a comparison of the relative energies between the different \ch{CsPbI3} material phases. Without any experimental data on the formation energies of the compounds, the different approximations were compared amongst themselves. The resulting formation energies are shown in Fig. \ref{energy_comparison}.

From the comparison of the geometries, we learn that the majority of density functional approximations accurately predicts the experimental unit cell geometries (to within 5\% volume), with PBE + DFT-D3(BJ) performing the best (to within 2\% volume). Two notable outliers in this comparison are LDA and PBE. However, the mismatch of LDA and PBE with the experimental geometries is of no surprise, since they are known for under- and overpredicting geometries due to a respective over- and underbinding~\cite{csonkaAssessingPerformanceRecent2009, zhangPerformanceVariousDensityfunctional2018}. Our geometry comparison demonstrates that the underbinding of PBE can be overcome with some refinements to the density functional approximation, i.e. the reparametrization of PBE for solids (PBEsol) or the explicit inclusion of dispersion interactions through a post-correction (DFT-D). The improvement is most apparent for the layered \ch{PbI2} compound. However, the refinement of PBE with DFT-D3 does not improve the result for all material geometries: the underbinding of PBE persists in the calculation of \ch{CsI} with PBE + DFT-D3. In contrast to LDA and PBE, the SCAN meta-GGA functional predicts geometries that are rather close to experiments. Although SCAN itself already accurately predicts geometries, the explicit inclusion of dispersive interactions, either through DFT-D methods or rVV10, results in predictions that agree slightly better with experiments, nevertheless the improvement is not as distinct as for PBE.

An inspection of the formation energies shows that they are sensitive to the choice of XC-functional, which has been reported for the \ch{CsPbI3} metal halide perovskite in literature before~\cite{suttonCubicOrthorhombicRevealing2018}. Despite the sensitivity to the choice of XC-functional, all density functional approximations predict a similar stability trend for the different \ch{CsPbI3} phases. When ordered from least to most stable, the phases are cubic, tetragonal, orthorhombic and yellow phase \ch{CsPbI3}, a trend that matches well with experiments~\cite{stoumposRenaissanceHalidePerovskites2015, marronnierAnharmonicityDisorderBlack2018}. Nonetheless, when the stability of \ch{CsPbI3} with respect to its precursors is concerned, PBE + DFT-D3 appears to be an outlier. It predicts all perovskite phases of \ch{CsPbI3} to be stable with respect to its precursors, whereas most approximations predict at least a part to be unstable to its precursors. This apparent overstabilization is of geometric origin. In the benchmark of the geometries, we found that PBE + DFT-D3 underbinds \ch{CsI}. Naturally, this results in a destabilization (higher energy) of this precursor. Thus, all the formation energies for this density functional approximation are overstabilized due to this erroneous reference structure.

When we take into account the results of both benchmarks, we judge the PBE + DFT-D3(BJ) approximation to be the most suitable for the generation of the reference data in the training set. Not only does it result in geometries that best resemble experimental geometries, it also predicts a phase stability for the phases of \ch{CsPbI3} that is in line with experimental observations with a reasonable computational cost. Therefore, we generate all reference data in our training set with the PBE + DFT-D3(BJ) density functional approximation.

\subsection*{Supplementary Note 3: Reversibility of perovskite phase transitions}

\begin{figure}[htbp!]
    \includegraphics[width=0.9\textwidth]{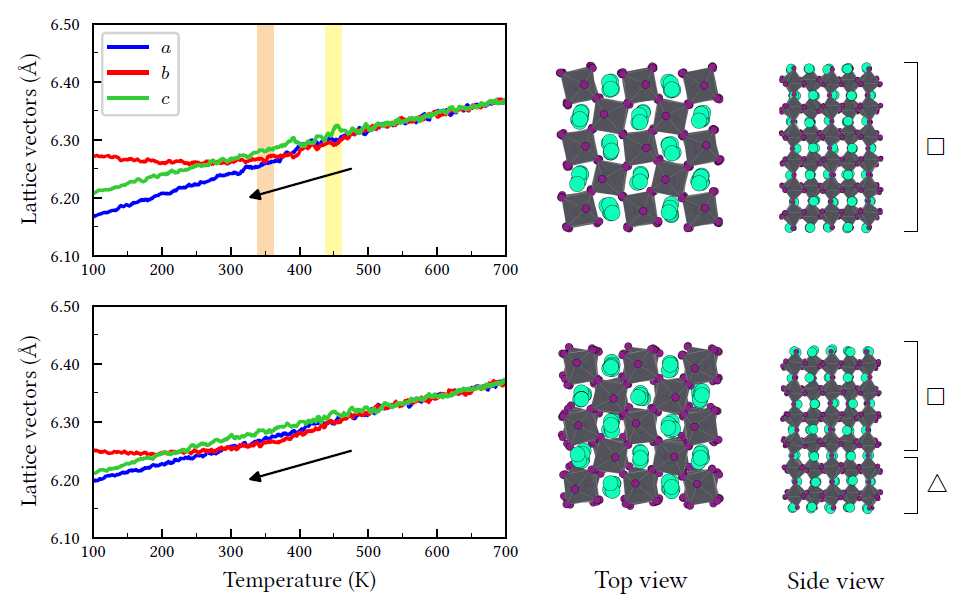}
    \caption{\label{reversibility} \textbf{Comparison of the reversibility of the \ch{CsPbI3} phase transitions.} A comparison of two types of ReaxFF cooling runs during which a model system with 864 formal units of \ch{CsPbI3} was cooled. The comparison shows a completely reversible (top panel) and partially reversible (bottom panel) cooling run. The structural snapshots shown for each type of cooling run demonstrate the formation of a single orthorhombic domain for the completely reversible cooling run and two differently oriented domains for the partially reversible cooling run.}
\end{figure}

To obtain atomic-scale insights into the incomplete reversibility of the phase transitions of \ch{CsPbI3}, we compare the phase diagrams and structures from a cooling run that exhibits complete reversibility to one that only shows a partial reversibility. The investigated structures from both simulations were obtained from the atomic positions of the configuration in the final frame of the simulation, which is a representation of the structure into which the system cools. The comparison of these two types of simulation runs in shown in Fig. \ref{reversibility}.

The two phase diagrams exhibit a distinct difference: in the partially reversible phase diagram the lattice vectors do not show an as distinct splitting of the lattice vectors at low temperatures as seen in a completely reversible phase diagram. This incomplete splitting of the lattice vectors is supported by the atomic structures. We note that during the completely reversible cooling run, the perovskite adopts an orthorhombic phase with a single orientation (denoted with the square), as demonstrated by the in phase rotation of the \ch{PbI6} octahedra in the top view. As such, the \ch{CsPbI3} here adopts a single orthorhombic domain. In contrast, for the partially reversible run, the top view of the atomic structures shows a staggered configuration of the \ch{PbI6} octahedra, that results from an out-of-phase rotation of the octahedra in consecutive perovskite layers. Thus, the perovskite adopts a structure composed of differently oriented orthorhombic domains (denoted with the square and triangle), a phenomenon also found for \ch{CsPbI3} in experiments where it is referred to as \textit{twinning}~\cite{bertolottiCoherentNanotwinsDynamic2017}. On the basis of the run-to-run variation, we hypothesize that whether or not we observe the formation of a single domain (complete reversibility), is dependent on the local structural fluctuations encountered during a cooling run.

\subsection*{Supplementary Note 4: Lattice vector analysis}

During the constant temperature simulations of the perovskite used in construction of the constant temperature phase diagram, the lattice vectors fluctuated considerably, making it difficult to uniquely determine the value of the lattice vectors at these these temperatures. To do so, we closely follow an approach proposed by Jinnouchi et al.~\cite{jinnouchiPhaseTransitionsHybrid2019}.

In this approach, we bin the time evolution of the pseudo-cubic lattice vectors, the result of which is consequently fit with Gaussian distributions to determine the value of the lattice vectors. The Gaussian distributions $\rho$ have the following functional form
\begin{equation}
    \rho \left( x \right) = A \exp \left[ - \frac{1}{2} \left( \frac{x - \mu}{\sigma} \right)^{2} \right],
\end{equation}
where $x$ is the binned lattice vector distribution, $A$ a scaling factor of the distribution, $\mu$ the equilibrium lattice vector and $\sigma$ the standard deviation of the lattice vector distribution.

First, all pseudo-cubic lattice vectors were binned individually, where each binned distribution was fit with their own Gaussian distribution. Each of the distributions has its own average value $\mu_{i}$ and standard deviation $\sigma_{i}$, with $i = 1, 2$ or $3$ indexing the different distributions. If the difference between the average value of the lattice vectors was larger than $\frac{1}{3} \sqrt{\sigma_{1}^{\, 2} + \sigma_{2}^{\, 2} + \sigma_{3}^{\, 2}}$, the unit cell was judged orthorhombic and consequently resulting in a unique value for all lattice vectors ($\mu_{1}, \mu_{2} \text{ and } \mu_{3}$). If not, the lattice vectors with a smaller difference were grouped, and two Gaussian distributions were fit ($i = 1, 2$). Now, in case the difference between the average values of these two distributions was larger than $\frac{1}{2} \sqrt{\sigma_{1}^{\, 2} + \sigma_{2}^{\, 2}}$, the unit cell was judged tetragonal, resulting in two values for the lattice vectors ($\mu_{1} \text{ and } \mu_{2}$). Else, the unit cell was judged as cubic, grouping all lattice vectors together, resulting in a single Gaussian function to fit ($i = 1$) and a single value for the lattice vectors ($\mu_{1}$).

\subsection*{Supplementary Note 5: Harmonic approximation for perovskite lattice}

For the derivation of the harmonic approximation of the probability distribution of the position of the \ch{I} atom in the system, we employ a polar coordinate system. The origin of this coordinate system is defined by the line that interconnects the two neighbouring \ch{Pb} atoms. We define the displacement away from this line, which we referred to as $\delta$ in the polar coordinates $r$ and $\phi$. In the harmonic approximation, we assume the iodine atom experiences a harmonic potential $U$ as
\begin{equation}
    U \left( r, \phi \right) = \frac{1}{2} k r^{2}
\end{equation}
where $k$ is an arbitrary force constant. The lack of any angular dependence on $\phi$ is a result of the isotropy of the system. By combining this potential with the Boltzmann distribution, we can construct a polar probability density $p$ as
\begin{equation}
    p \left( r, \phi \right) \propto \exp \left( - \frac{U}{k_{\text{B}} T} \right) = \exp \left(- \frac{r^{2}}{2 \sigma^{2}} \right),
\end{equation}
where in the final step we filled out the harmonic potential and rewrote the expression so that it mimics a normal distribution with $\sigma = \sqrt{\frac{k_{\text{B}} T}{k}}$. The probability density can be used to calculate $P \left( r \right) dr$, the probability of finding the iodine atom in the interval from $r$ to $r + dr$, through an integration of the polar angle $\phi$, as
\begin{equation}
    P \left( r \right) dr = \left[ \int^{2 \pi}_{0} p \left( r, \phi \right) r \, d\phi \right] dr \propto r \exp \left(- \frac{r^{2}}{2 \sigma^{2}} \right) dr.
\end{equation}
The probability distribution for $r$ that we are then left with is
\begin{equation}
    P \left( r \right) = A \, r \exp \left(- \frac{r^{2}}{2 \sigma^{2}} \right),
\end{equation}
in which both $A$ and $\sigma$ are parameters that can be used to fit the harmonic model to the simulation data.

\clearpage

\section*{Supplemenatary Tables}

\subsection*{Supplementary Table 1: Self-diffusion coefficients}

\begin{table}[htbp!]
    \caption{Self-diffusion coefficients of iodine point defects two iodine interstitials (I\textsubscript{I}) and two iodine vacancies (V\textsubscript{I}) obtained with ReaxFF simulations of \ch{CsPbI3} with the average values in brackets.}
    \begin{ruledtabular}
    \begin{tabular}{ccc}
        T (K)       & $D_{\text{I}_{\ch{I}}} \left( \times 10^{-9} \SI{}{\square\cm\per\s} \right)$  & $D_{\text{V}_{\ch{I}}} \left( \times 10^{-9} \SI{}{\square\cm\per\s} \right)$ \\
        \colrule
        450         & 1.76, 1.88, 2.48; \textbf{[2.04]}                                             & 2.31, 2.62; \textbf{[2.47]} \\
        500         & 3.72, 5.15, 5.35; \textbf{[4.74]}                                             & 3.13, 3.66, 4.02, 4.88, 5.02, 5.65; \textbf{[4.39]} \\
        550         & 7.66, 9.32, 10.92; \textbf{[9.30]}                                            & 4.03, 5.73, 7.84; \textbf{[5.87]} \\
        600         & 13.47, 13.85, 14.57; \textbf{[13.97]}                                         & 7.95, 9.77; \textbf{[8.86]} \\
        650         & 18.38, 18.97, 20.26; \textbf{[19.21]}                                         & - \\
        700         & 26.69, 27.87, 30.45; \textbf{[28.34]}                                         & - \\
    \end{tabular}
    \end{ruledtabular}
\end{table}

\clearpage

\section*{Supplementary Figures}

\subsection*{Supplementary Figure 1: Data smoothing}

\begin{figure}[htbp!]
    \includegraphics[width=0.9\textwidth]{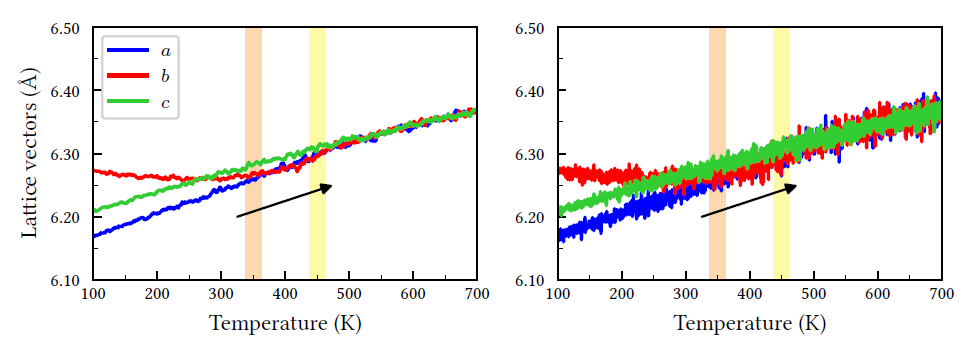}
    \caption{\textbf{Effect of the data smoothing.} An overview of the phase diagrams obtained with a without data smoothing during a gradual heating of a $6 \times 6 \times 6$-supercell of orthorhombic \ch{CsPbI3} (864 formal units). A \SI{10}{\ps} running average is used to smooth the data (left panel) whereas the other panel (right panel) present the raw data. The orange and yellow bars, respectively, indicate the phase transition temperatures for the orthorhombic to tetragonal and tetragonal to cubic phase transitions from the ReaxFF simulations of \ch{CsPbI3}. In all figures the pseudo-cubic lattice vectors, $a$, $b$ and $c$, of \ch{CsPbI3} are used.}
\end{figure}

\clearpage

\subsection*{Supplementary Figure 2: Heating rates}

\begin{figure}[htbp!]
    \includegraphics[width=0.9\textwidth]{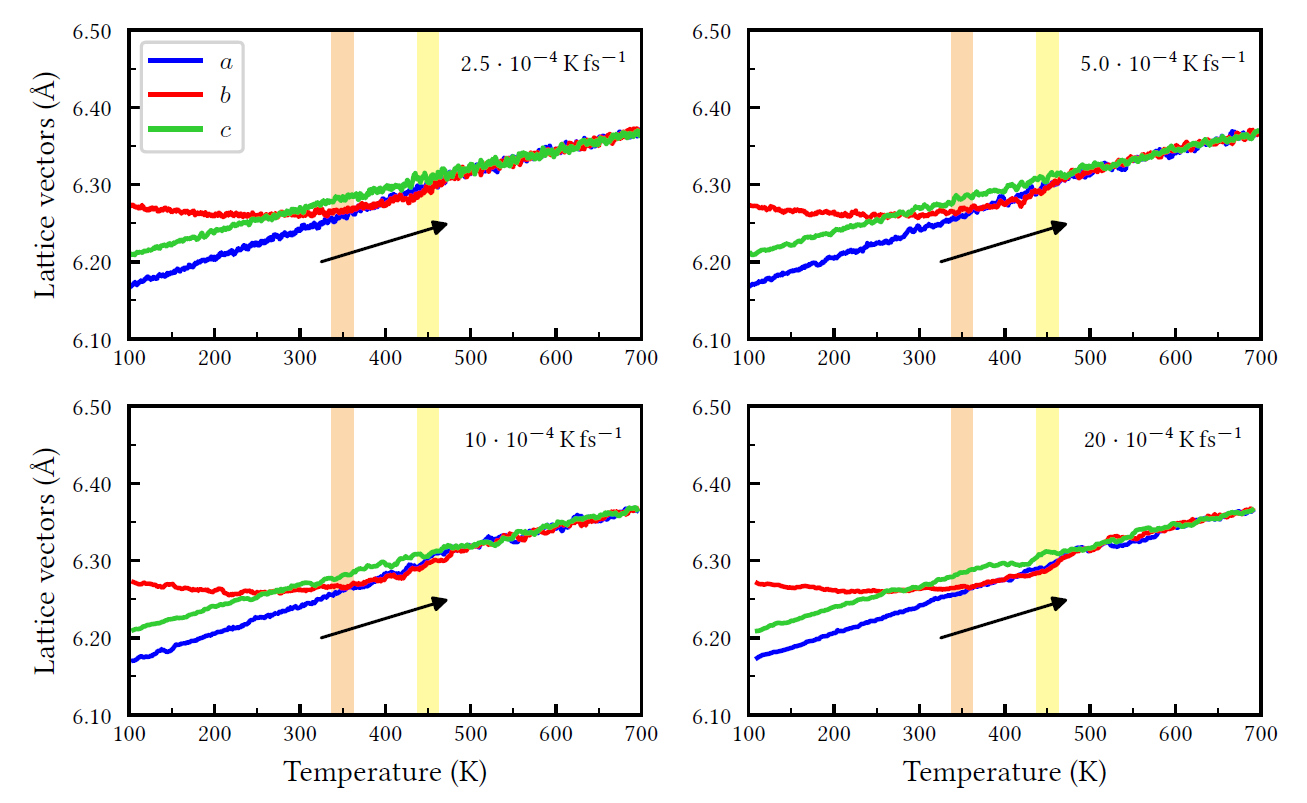}
    \caption{\textbf{Comparison of different heating rates.} An overview of the phase diagrams obtained for a variety of temperature ramp rates during the gradual heating of a $6 \times 6 \times 6$-supercell of orthorhombic \ch{CsPbI3} (864 formal units). The orange and yellow bars, respectively, indicate the phase transition temperatures for the orthorhombic to tetragonal and tetragonal to cubic phase transitions from the ReaxFF simulations of \ch{CsPbI3}. In all figures the pseudo-cubic lattice vectors, $a$, $b$ and $c$, of \ch{CsPbI3} are used.}
\end{figure}

\clearpage

\subsection*{Supplementary Figure 3: Thermostat and barostat type}

\begin{figure}[htbp!]
    \includegraphics[width=0.9\textwidth]{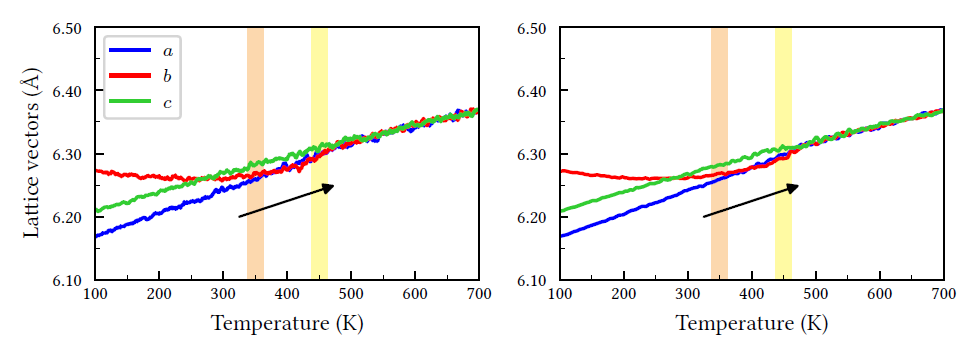}
    \caption{\textbf{Comparison of different temperature and pressure control mechanisms.} An overview of the phase diagrams obtained using different temperature and pressure control mechanisms during the gradual heating of a $6 \times 6 \times 6$-supercell of orthorhombic \ch{CsPbI3} (864 formal units). The used combinations are a NHC-thermostat and MTK-barostat (left panel) and a Berendsen thermostat and Berendsen barostat (right panel). The orange and yellow bars, respectively, indicate the phase transition temperatures for the orthorhombic to tetragonal and tetragonal to cubic phase transitions from the ReaxFF simulations of \ch{CsPbI3}. In all figures the pseudo-cubic lattice vectors, $a$, $b$ and $c$, of \ch{CsPbI3} are used.}
\end{figure}

\clearpage

\subsection*{Supplementary Figure 4: Mean square displacements}

\begin{figure}[htbp!]
    \includegraphics[width=0.9\textwidth]{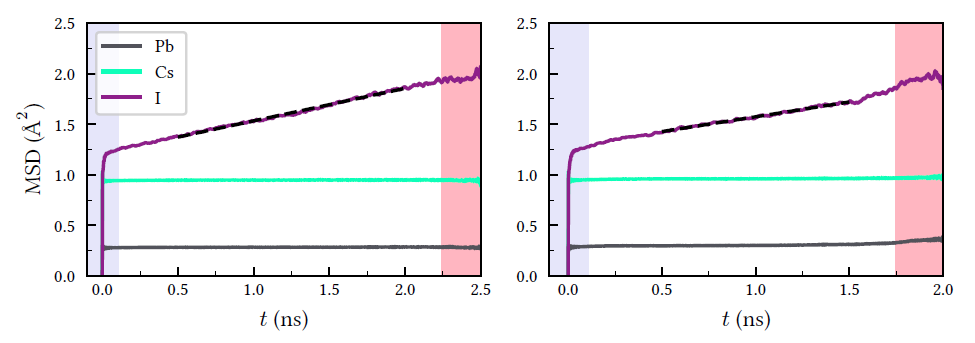}
    \caption{\textbf{An overview of the mean square displacements of defective \ch{CsPbI3}.} Examples of the mean square displacement (MSD) of the chemical species in defective \ch{CsPbI3} from ReaxFF simulations. The panels show the results of simulations at \SI{500}{\K} for a model system containing two iodine interstitials (left panel) and two iodine vacancies (right panel). A linear fit (dashed line) is made to the MSD to determine the diffusion coefficients.}
\end{figure}

\clearpage

\bibliography{supplementary}% Produces the bibliography via BibTeX.